\documentstyle[11pt]{article}
\def\agt{
\mathrel{\raise.3ex\hbox{$>$}\mkern-14mu\lower0.6ex\hbox{$\sim$}}
}
\def\alt{
\mathrel{\raise.3ex\hbox{$<$}\mkern-14mu\lower0.6ex\hbox{$\sim$}}
}
\begin{document}

\bibliographystyle{unsrt}    

\Large
{\centerline{\bf Probing Black Holes and Relativistic Stars }} 

\vspace{2mm}

\Large
{\centerline{\bf with Gravitational Waves}}
\small\normalsize

\vspace{7mm}

\centerline {\bf Kip S. Thorne}
\small\normalsize

\vspace{3mm}

\centerline {California Institute of Technology, Pasadena, CA 91125 USA}

\vspace{4mm}

\begin{abstract}
In the coming decade,
gravitational waves will convert the study of general relativistic aspects of
black holes
and stars from a largely theoretical enterprise to a highly
interactive, observational/theoretical one.  
For example, gravitational-wave observations should enable us to
observationally map the spacetime geometries around quiescient black holes,
study quantitatively the highly nonlinear vibrations of curved spacetime
in black-hole collisions, probe the structures of neutron stars and their
equation of state, search for exotic types of general relativistic objects such
as boson stars, soliton stars, and naked singularities, and probe aspects of
general relativity that have never yet been seen such as the gravitational
fields of gravitons and the influence of gravitational-wave tails on radiation
reaction.
\end{abstract}

\vspace{1mm}

\section{Introduction}\label{introduction} 

Subrahmanyan Chandrasekhar and I entered the field of relativistic astrophysics
at the same time, in the early 1960s---I as a green graduate student at
Princeton; Chandra as an estabilished and famous researcher at the University of
Chicago.  Over the decades of the 60's, 70's and 80's, and into the 90's,
Chandra, I, and our friends and colleagues had the great pleasure of exploring
general relativity's predictions about the properties of black holes and
relativistic stars.  Throughout these explorations Chandra was an inspiration
for us all.  

When we began, there was no observational evidence that black holes or
relativistic stars exist in the Universe, much less that they play important 
roles.  However, 
in parallel with our theoretical studies, astronomers discovered pulsars
and quickly deduced they are spinning neutron stars in which relativistic
effects should be strong;
astronomers also discovered quasars,
and gradually, over three decades' time, came to understand that they
are powered by supermassive black holes; astronomers discovered compact 
X-ray sources and quickly deduced they are binary systems in which gas 
accretes from a normal star onto a stellar-mass black-hole companion or 
neutron-star companion; and astronomers discovered gamma ray bursts, and after
nearly three decades of puzzlement, have concluded they are probably produced 
by the final merger of a neutron-star /neutron-star binary or a neutron-star
/ black-hole binary.

Despite this growing richness of astrophysical phenomena in which black holes
and neutron stars play major roles, those of us who use general
relativity to predict the properties of these objects have been frustrated:
in the rich astronomical data there as yet is little evidence for the
holes' and stars' spacetime warpage, which is so central to our theoretical 
studies.  If we had to rely solely on observations and not at all on theory, 
we could still argue, in 1997, that
a black hole is a flat-spacetime, Newtonian phenomenon and neutron stars are
un-influenced by general relativistic effects.

Why this frustration?  Perhaps because spacetime warpage cannot, itself,
produce the only kinds of radiation that astronomers now have at their
disposal: electromagnetic waves, neutrinos, and cosmic rays.  To explore 
spacetime
warpage in detail may well require using, instead, the only kind of radiation
that such warpage can produce: radiation made of spacetime
warpage---gravitational waves.  

In this article, I shall describe the prospects for using gravitational waves
to probe the warpage of spacetime around black holes and relativistic stars,
and to search for new types of general relativistic objects, for which there as
yet is no observational evidence.  And I shall describe how the challenge of
developing data analysis algorithms for gravitational-wave detectors is already
driving the theory of black holes and relativistic stars just as hard as
the theory is driving the wave-detection efforts.  Already, several years
before the full-scale detectors go into operation, the challenge of
transforming general relativistic astrophysics into an observational science
has transformed the nature of our theoretical enterprise.  At last, after 35
years of only weak coupling to observation, those of us studying general
relativistic aspects of black holes and stars have become tightly
coupled to the observational/experimental enterprise.

\section{Gravitational Waves}\label{gravwaves}

A gravitational wave is a ripple of warpage (curvature) in the ``fabric'' of
spacetime.  According to general relativity, gravitational 
waves are produced by 
the dynamical spacetime warpage of distant
astrophysical systems, and they travel outward from their sources and through 
the Universe at the speed of light, becoming very weak by the time they 
reach the Earth.  
Einstein discovered gravitational waves as a prediction of his
general relativity theory in 1916, but only in the late 1950s did the 
technology of high-precision measurement become good enough to justify 
an effort to construct detectors for the waves.

Gravitational-wave detectors and detection techniques have now been under 
development for nearly 40 years, building on foundations laid by
Joseph Weber \cite{weber}, Rainer Weiss \cite{weiss}, and others.
These efforts have led to promising sensitivities in four frequency 
bands, and theoretical studies have identified plausible sources in
each band:  
\begin{itemize}
\item
The Extremely Low Frequency Band (ELF), $10^{-15}$ to $10^{-18}$ Hz, in which
the measured anisotropy of the cosmic microwave background radiation places
strong limits on gravitational wave strengths---and may, in fact, have
detected waves \cite{turner,grishchuk96}.  The only waves expected in this band
are relics of the big bang, a subject beyond the scope of this article.  (For
some details and references see \cite{turner,grishchuk96,veneziano95} and 
references cited therein.)
\item
The Very Low Frequency Band (VLF), $10^{-7}$ to $10^{-9}$ Hz, in which Joseph
Taylor and others have achieved remarkable gravity-wave sensitivities by the
timing of millisecond pulsars \cite{kaspi_taylor}.  The only expected strong
sources in this band are processes in the very early universe---the big bang,
phase transitions of the vacuum states of quantum fields, and vibrating or
colliding defects in the structure of spacetime, such as monopoles,
cosmic strings, domain walls, textures, and combinations thereof 
\cite{zeldovich_strings,vilenkin_strings,turner_kosowsky,martin_vilenkin}.  
These sources are
also beyond the scope of this article. 
\item
The Low-Frequency Band (LF), $10^{-4}$ to 1 Hz, in which will operate the Laser
Interferometer Space Antenna, LISA; see Sec.\ \ref{lisa} below.  This is 
the band of massive black holes 
($M\sim 1000$ to $10^8 M_\odot$) in the distant universe,
and of other hypothetical massive exotic objects (naked singularities, soliton
stars), as well as of binary stars (ordinary, white dwarf, 
neutron star, and black
hole) in our galaxy. Early universe processes 
should also have produced waves at these frequencies, as in the ELF and 
VLF bands.
\item
The High-Frequency Band (HF), $1$ to $10^4$Hz, in which operate earth-based
gravitational-wave detectors such as LIGO; 
see Secs.\ \ref{gbint}--\ref{narrowband} below.  This 
is the band of stellar-mass black holes ($M \sim 1$ to $1000M_\odot$) and 
of other conceivable stellar-mass exotic objects 
(naked singularities and boson stars) in the distant universe,
as well as of supernovae, pulsars, and coalescing and colliding neutron stars.
Early universe processes should also have produced waves 
at these frequencies, as in the ELF, VLF, and LF bands.
\end{itemize}

In this article I shall focus on the HF and LF bands, because these
are the ones in which we can expect to study black holes and
relativistic stars.

One aspect of a gravitational wave's spacetime warpage---the only aspect 
relevant to earth-based detectors---is 
an oscillatory ``stretching and squeezing'' of space.  This
stretch and squeeze is described, in general relativity theory, by two 
dimensionless gravitational wave fields 
$h_+$ and $h_\times$
(the ``strains of space'') that are
associated with the wave's two linear polarizations,
conventionally called ``plus'' ($+$) and ``cross'' ($\times$).  The fields 
$h_+$ and $h_\times$, technically speaking, are the double time integrals
of space-time-space-time components of the Riemann curvature tensor;
and they propagate through spacetime at the speed of light.  The inertia of any
small piece of an object tries to keep it at rest in, or moving at
constant speed through, the piece of space in which it resides; so as $h_+$ and
$h_\times$ stretch and squeeze space, inertia stretches and squeezes objects 
that reside
in that space.  This stretch and squeeze is analogous
to the tidal gravitational stretch and squeeze exerted on the Earth by the 
Moon, and thus the associated gravitational-wave force 
is referred to as a ``tidal'' force.   

\begin{figure}
\vskip 9.pc
\special{hscale=65 vscale=65 hoffset=42 voffset=-13
psfile=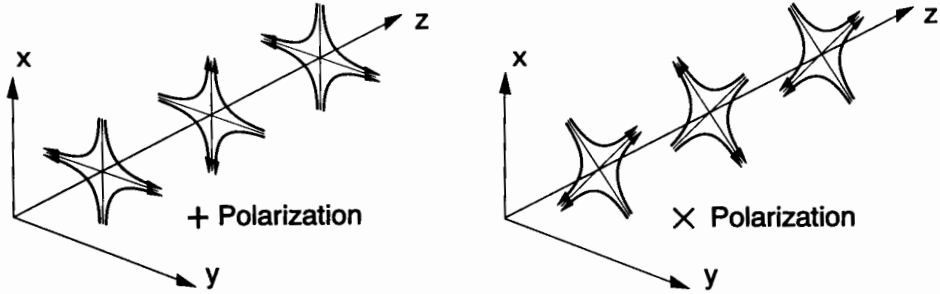}
\caption{The lines of force associated with the two polarizations of
a gravitational wave.  (From Ref. \protect\cite{ligoscience}.)
}
\label{fig:forcelines}
\end{figure}

If an object is small
compared to the waves' wavelength (as is the case for ground-based
detectors), then relative to the
object's center, the waves exert tidal forces  with the 
quadrupolar patterns shown in Fig.~\ref{fig:forcelines}.
The names ``plus'' and ``cross'' are derived from the orientations of the
axes that characterize the force patterns \cite{300yrs}. 

The strengths of the waves from a gravitational-wave source can be 
estimated using the
``Newtonian/quadrupole'' approximation to the Einstein field equations.
This approximation says that $h\simeq (G/c^4)\ddot Q/r$, where
$\ddot Q$ is the second time derivative of the source's quadrupole
moment, $r$ is the distance of the source from Earth (and $G$ and $c$
are Newton's gravitation constant and the speed of light).  
The strongest sources will be highly nonspherical and thus will
have $Q\simeq ML^2$, where $M$ is their mass and $L$ their size, and
correspondingly will have $\ddot Q \simeq 2Mv^2 \simeq 4 E_{\rm
kin}^{\rm ns}$, where $v$ is their internal velocity and
$E_{\rm kin}^{\rm ns}$ is the nonspherical part of their
internal kinetic energy.  This provides us with the estimate
\begin{equation}
h\sim {1\over c^2}{4G(E_{\rm kin}^{\rm ns}/c^2) \over r}\;;
\label{hom}
\end{equation}
i.e., $h$ is about 4 times the gravitational potential produced at Earth by
the mass-equivalent of the source's nonspherical, internal kinetic 
energy---made dimensionless by dividing by $c^2$.  Thus, in order to
radiate strongly, the source must have a very large, nonspherical,
internal kinetic energy.

The best known way to achieve a huge internal kinetic energy is via
gravity; and by energy conservation (or the virial theorem), any
gravitationally-induced kinetic energy must be of order the source's 
gravitational potential energy.  A huge potential energy, in turn, 
requires that the source be very compact, not much larger than its own 
gravitational radius.  Thus, the strongest gravity-wave sources must be
highly compact, dynamical concentrations of large amounts of mass (e.g.,
colliding and coalescing black holes and neutron stars).

Such sources cannot remain highly dynamical for long; their motions will
be stopped by energy loss to gravitational waves and/or the formation of
an all-encompassing black hole.  Thus, the strongest sources should be
transient.  Moreover, they should be very rare---so rare that to see a
reasonable event rate will require reaching out through a substantial
fraction of the Universe.  Thus, just as the strongest radio waves 
arriving at Earth 
tend to be extragalactic, so also the strongest gravitational waves are 
likely to be extragalactic.  

For highly compact, dynamical objects that radiate in the high-frequency band,
e.g.\ colliding and coalescing neutron stars and stellar-mass black
holes, the internal, nonspherical kinetic energy 
$E_{\rm kin}^{\rm ns}/c^2$ is of order the mass of
the Sun; and, correspondingly, Eq.\ (\ref{hom}) gives $h\sim 10^{-22}$
for such sources at the Hubble 
distance (3000 Mpc, i.e., $10^{10}$ light years); 
$h\sim 10^{-21}$ at 200 Mpc (a best-guess distance for several
neutron-star coalescences per year; see Section \ref{nsnsinspiral}), 
$h\sim 10^{-20}$ at the
Virgo cluster of galaxies (15 Mpc); and $h\sim 10^{-17}$ in the outer
reaches of our own Milky Way galaxy (20 kpc).  These numbers set the
scale of sensitivities that ground-based interferometers seek to achieve:
$h \sim 10^{-21}$ to $10^{-22}$.  

\section{Gravitational Wave Detectors in the High and Low Frequency Bands}

\subsection{Ground-Based Laser Interferometers}\label{gbint}

\begin{figure}
\center
\vskip 13.2pc
\special{hscale=70 vscale=70 hoffset=45 voffset=-5
psfile=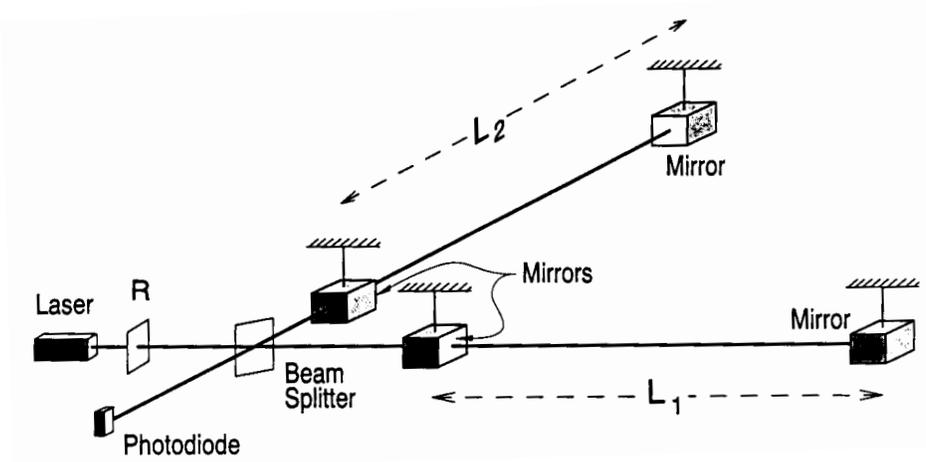}
\caption{Schematic diagram of a laser interferometer
gravitational wave detector.  (From Ref.\ \protect\cite{ligoscience}.)
}
\label{fig:interferometer}
\end{figure}

The most promising and versatile type of gravitational-wave detector in the
high-frequency band, $1$ to $10^4$Hz, is a
laser interferometer gravitational wave detector (``interferometer'' for
short).  Such an interferometer consists of four mirror-endowed
masses that hang from 
vibration-isolated 
supports as shown in Fig.\ \ref{fig:interferometer}, and the indicated 
optical system for 
monitoring the separations between
the masses \cite{300yrs,ligoscience}. Two masses are near 
each other, at the corner of an ``L'', 
and one mass is at the end of each of the L's long arms.  The arm
lengths are nearly equal, $L_1 \simeq L_2 = L$.  When a gravitational
wave, with frequencies high compared to the masses' $\sim 1$ Hz pendulum
frequency, passes through the detector, it pushes the masses back and
forth relative to each other as though they were free from their
suspension wires, thereby changing the arm-length difference,
$\Delta L \equiv L_1-L_2$.  That change is
monitored by laser interferometry in such a way that the variations in
the output of the
photodiode (the interferometer's output) are directly proportional to
$\Delta L(t)$.

If the waves are coming from overhead or
underfoot
and the axes of the $+$ polarization coincide with the arms'
directions, then it is the waves' $+$ polarization that drives the masses, and
the detector's strain $\Delta L(t) / L$ is equal to the waves' strain of space
$h_+(t)$.  More generally, 
the interferometer's output is a linear combination of the two 
wave fields: 
\begin{equation}
{\Delta L(t)\over L} = F_+h_+(t) + F_\times h_\times (t) \equiv h(t)\;.
\label{dll}
\end{equation}
The coefficients $F_+$ and $F_\times$ are of order unity and depend in a
quadrupolar manner on 
the direction to the source and the orientation of the detector \cite{300yrs}.
The combination $h(t)$ of the two $h$'s is called the
gravitational-wave strain that acts on the detector; and the time
evolutions of $h(t)$, $h_+(t)$, and $h_\times(t)$ are sometimes called
{\it waveforms}.

When one examines the technology of laser interferometry, one sees good
prospects to achieve measurement accuracies $\Delta L \sim 10^{-16}$ cm
(1/1000 the diameter of the nucleus of an atom)---and $\Delta L =
8\times10^{-16}$ has actually been achieved in a prototype interferometer at
Caltech \cite{40mspectrum}.  With $\Delta L \sim 10^{-16}$cm,
an interferometer must have an arm length $L = \Delta L/h \sim 1$ to 10
km in order to achieve the desired wave sensitivities, $10^{-21}$ to
$10^{-22}$.  This sets the scale of the interferometers that are now
under construction.

\subsection{LIGO, VIRGO, and the International Network of Gravitational Wave
Detectors}
\label{network}

Interferometers are plagued by non-Gaussian noise, e.g.\ due
to sudden strain releases in the wires that suspend the masses.  This
noise prevents a single interferometer, by itself, from detecting with
confidence short-duration gravitational-wave bursts (though it may
be possible for a single interferometer to search for the periodic
waves from known pulsars).   The non-Gaussian noise can be removed
by cross correlating two, or preferably three or more, interferometers 
that are networked together at widely separated sites. 

The technology and techniques for such interferometers have been under
development for 25 years, and plans for km-scale 
interferometers have been developed over the past 15 years.  An
international network consisting of three km-scale interferometers at
three widely separated sites is now under 
construction. It includes two sites of 
the American LIGO Project (``Laser Interferometer Gravitational Wave
Observatory'') \cite{ligoscience}, and one site of the French/Italian VIRGO 
Project (named after the Virgo cluster of galaxies) \cite{virgo}.  

LIGO will consist of two 
vacuum facilities with 4-kilometer-long arms, one in Hanford, Washington 
(in the northwestern United States)
and the other in Livingston, Louisiana (in the southeastern United States).  
These facilities are designed to house
many successive generations of interferometers without the necessity of
any major facilities upgrade; and after a planned
future expansion,
they will be able to house several interferometers at once, each
with a different optical configuration optimized for a different type of
wave (e.g., broad-band burst, or narrow-band periodic wave, or
stochastic wave).  

The LIGO facilities 
are being constructed by a team of about 80 physicists and engineers at
Caltech and MIT, led by Barry Barish (the PI), Gary Sanders (the
Project Manager), Albert Lazzarini, Rai Weiss, Stan Whitcomb, and
Robbie Vogt (who directed the project during the pre-construction phase).  
This Caltech/MIT team, together with researchers from several other 
universities, is developing
LIGO's first interferometers and their data analysis system.  
Other research groups from many 
universities are contributing to R\&D for {\it enhancements} of the first
interferometers, or are computing theoretical waveforms
for use in data analysis, or are developing data analysis techniques for
future interferometers.  
These groups are linked together in a {\it LIGO Scientific
Collaboration} and by an 
organization called the {\it LIGO Research Community}.
For further details, see the LIGO World Wide Web Site, 
http://www.ligo.caltech.edu/. 

The VIRGO Project is building one vacuum facility in Pisa, Italy, with 
3-kilometer-long arms.  This facility and its first interferometers are
a collaboration of more than a hundred physicists and engineers
at the INFN (Frascati, Napoli, Perugia, Pisa), LAL (Orsay), LAPP
(Annecy), LOA (Palaiseau), IPN (Lyon), ESPCI (Paris), and the University
of Illinois (Urbana), under the leadership of Alain Brillet and
Adalberto Giazotto.  

\begin{figure}
\center
\vskip23.8pc
\special{hscale=65 vscale=65 hoffset=40 voffset=-5
psfile=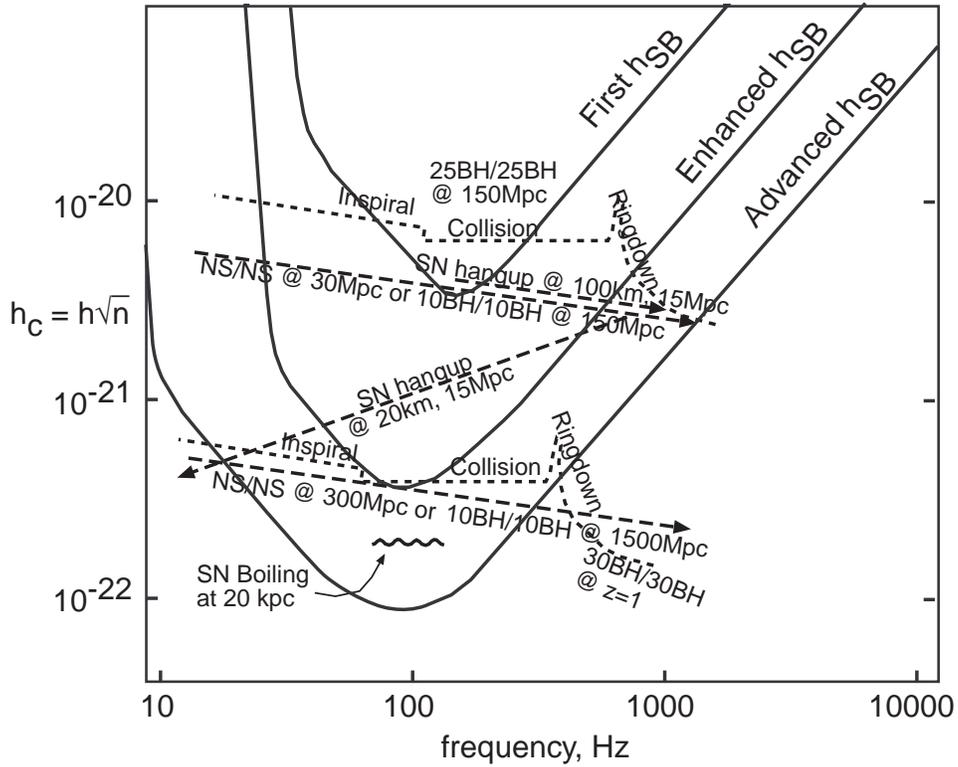}
\caption{LIGO's projected broad-band noise sensitivity
to bursts $h_{\rm SB}$ (Refs.\ 
\protect\cite{ligoscience,RandDproposal}) compared with the 
characteristic amplitudes $h_c$ 
of the waves from several hypothesized sources.  The
signal to noise ratios are 
$\protect\sqrt 2$ higher than in Ref.\ \protect\cite{ligoscience}
because of a factor 2 error in Eq.~(29) of Ref.\ \protect\cite{300yrs}.  
}
\label{fig:ligosources}
\end{figure}

The LIGO and VIRGO facilities are scheduled for completion
at the end of the 1990's, and their first gravitational-wave searches will
be performed in 2001 or 2002.  Figure \ref{fig:ligosources} shows the 
design sensitivities for LIGO's {\it first interferometers} (ca.\ 2001)
\cite{ligoscience} and for {\it enhanced versions} 
of those 
interferometers (which are expected to be operating five years or so later)
\cite{RandDproposal}, along with a benchmark sensitivity goal for 
subsequent, more {\it advanced interferometers} 
\cite{ligoscience,RandDproposal}.  

For each type of  
interferometer, the quantity shown is the ``sensitivity to bursts'' that come
from a random direction, $h_{\rm SB}(f)$ \cite{ligoscience}.  
This $h_{\rm SB}$ is about 5 times
worse than the rms noise level in a bandwidth $\Delta f \simeq f$
for waves with a random direction and polarization, and about $5\sqrt 5 
\simeq 11$ times worse 
than the the rms noise level $h_{\rm rms}$  for optimally 
directed and polarized waves.   (In much of the literature, the quantity
plotted is $h_{\rm rms} \simeq h_{\rm SB}/11$.)  Along the right-hand branch of
each sensitivity curve (above 100 or 200 Hz), the interferometer's 
dominant noise is due to photon counting statistics (``shot noise''); along the
middle branch (10 or 30 Hz to 100 to 200 Hz), the dominant noise is random
fluctuations of thermal energy in the test masses and their suspensions; along
the steep left-hand branch, the dominant 
noise is seismic vibrations creeping through
the interferometers' seismic isolation system.

The interferometer sensitivity $h_{\rm SB}$ is to be compared with the
``characteristic amplitude'' $h_c(f) = h \sqrt{n}$ of the waves from a source;
here $h$ is the waves' amplitude when they have frequency $f$, and $n$ is the
number of cycles the waves spend in a bandwidth $\Delta f \simeq f$ near
frequency $f$ \cite{300yrs,ligoscience}.  Any 
source with $h_c > h_{\rm SB}$ should be detectable with 
high confidence, even if it arrives only once per year.  

Figure
\ref{fig:ligosources} shows the estimated or computed characteristic 
amplitudes $h_c$ for several sources that will be discussed in detail later in
this article.  Among these sources are
binary systems made of 1.4$M_\odot$ neutron stars (``NS'') and binaries made
of 10, 25, and 30 $M_\odot$ black holes (``BH''), which spiral together 
and collide under the driving force
of gravitational radiation reaction.  As the bodies spiral inward, their
waves sweep upward in frequency (rightward across the figure along the dashed
lines).  From the figure we see that LIGO's first interferometers should be
able to detect waves from the inspiral of a NS/NS binary out to a distance 
of 30Mpc (90 million light years) and from the final collision and merger of a 
$25 M_\odot/25M_\odot$ BH/BH binary out to about 
300Mpc.  Comparison with estimated event rates (Secs.\ \ref{nsnsinspiral} 
and \ref{bhbhstrength} below) suggests,
with considerable confidence, that
the first wave detections will be achieved by the time the enhanced sensitivity
is reached and possibly as soon as the first-interferometers' searches.

LIGO alone, with its two sites which have parallel arms, will be able to
detect an incoming gravitational wave, measure one of its two waveforms,
and (from the time delay between the two sites) locate its source to within a
$\sim 1^{\rm o}$ wide annulus on the sky.
LIGO and VIRGO together, operating as a {\it coordinated international
network}, will be able to locate the source 
(via time delays plus the interferometers' beam patterns) 
to within a 2-dimensional error box with size
between several tens of arcminutes and several degrees, depending on
the source direction and on
the amount of high-frequency structure in the waveforms. 
They will also be able to monitor both waveforms $h_+(t)$ and
$h_\times(t)$ (except for frequency components above about 1kHz
and below about 10 Hz, where the interferometers' noise becomes severe). 

A British/German group is constructing a 600-meter interferometer called 
GEO 600 near Hanover Germany \cite{geo}, and Japanese groups, a 300-meter 
interferometer called TAMA near Tokyo \cite{tama}.  
GEO600 may be a significant player 
in the interferometric network in its early years (by virtue of cleverness and
speed of construction), but because of its short arms it cannot compete
in the long run.  GEO600 and TAMA will both be important development
centers and testbeds for interferometer techniques and technology, and 
in due course they may give rise to kilometer-scale 
interferometers like LIGO and VIRGO, which could significantly enhance the
network's all-sky coverage and ability to extract information from the waves.

\subsection{Narrow-Band, High-Frequency Detectors: Interferometers and 
Resonant-Mass Antennas}
\label{narrowband}

At frequencies $f\agt500$Hz, the interferometers' photon shot noise becomes a
serious obstacle to wave detection.  However, narrow-band detectors specially
optimized for kHz frequencies show considerable promise.  These include 
interferometers with specialized optical configurations 
(``signal recycled interferometers'' \cite{meers} and ``resonant sideband
extraction interferometers'' \cite{mizuno}), 
and large spherical or truncated icosahedral resonant-mass detectors 
(e.g., the American TIGA \cite{tiga}, Dutch GRAIL \cite{grail} and Brazilian
OMNI-1 Projects)
that are future variants of Joseph Weber's original ``bar'' detector
\cite{weber} and of currently operating bars in Italy (AURIGA, Explorer 
and Nautilus), 
Australia (NIOBE), and America (ALLEGRO) \cite{current_bars}.  
Developmental work for these narrow-band detectors is underway
at a number of centers around the world.

\subsection{Low-Frequency Detectors---The Laser Interferometer Space 
Antenna (LISA)}
\label{lisa}

The {\it Laser Interferometer Space Antenna} (LISA) \cite{cornerstone} is 
the most promising detector for gravitational waves in the low-frequency band,
$10^{-4}$--$1$ Hz (10,000 times lower than the LIGO/VIRGO high-frequency band).

LISA was originally conceived (under a different name) by Peter Bender of the
University of Colorado, and is currently being developed by an international 
team 
led by Karsten Danzmann of
the University of Hanover (Germany) and James Hough of Glasgow University (UK).
The European Space Agency tentatively plans to fly it sometime in the
2014--2018 time frame as part of ESA's Horizon 2000+ Program of large space
missions.  With NASA participation (which is under study), 
the flight could be much sooner.

\begin{figure}
\vskip15.8pc
\special{hscale=70 vscale=70 hoffset=68 voffset=7
psfile=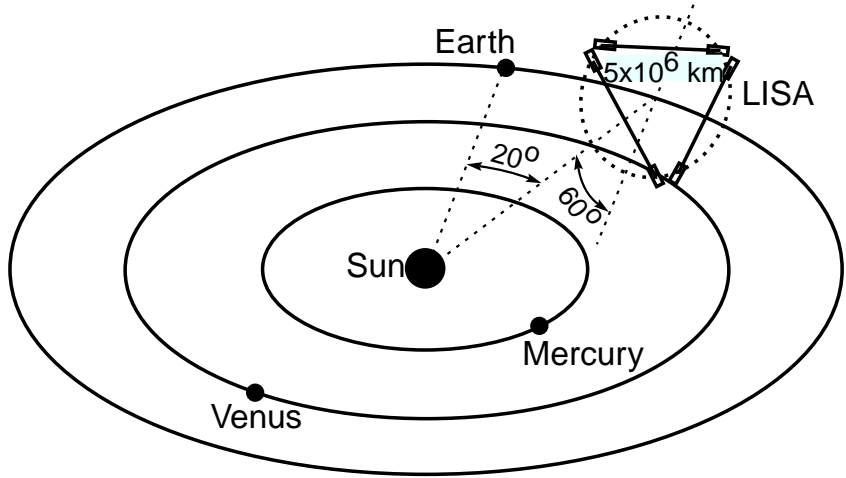}
\caption{LISA's orbital configuration, with LISA magnified in arm length
by a factor $\sim 10$ relative to the solar system.
}
\label{fig:lisa_orbit}
\end{figure}

As presently conceived \cite{cornerstone}, 
LISA will consist of six compact, drag-free
spacecraft (i.e. spacecraft that are shielded from buffeting by solar
wind and radiation pressure, and that thus move very nearly on geodesics of
spacetime).  All six spacecraft would be launched simultaneously in a 
single Ariane rocket. They
would be placed into the same heliocentric orbit as the Earth
occupies, but would follow 20$^{\rm o}$ behind the Earth; cf.\ Fig.\ 
\ref{fig:lisa_orbit}.  The spacecraft would fly in pairs, with each pair
at the vertex of an equilateral triangle that is inclined at an angle of 
60$^{\rm o}$ to the Earth's orbital plane. The triangle's arm length would be 5
million km ($10^6$ times longer than LIGO's arms!).  The six spacecraft would
track each other optically, using one-Watt YAG laser beams.  Because of 
diffraction
losses over the $5\times10^6$km arm length, it is not feasible to
reflect the beams back and forth between mirrors as is done with LIGO.
Instead, each spacecraft would have its own laser; and the lasers would be
phase locked to each other, thereby achieving the same kind of
phase-coherent out-and-back light travel as LIGO achieves with mirrors.
The six-laser, six-spacecraft configuration thereby would function as three,
partially independent and partially redundant, 
gravitational-wave interferometers.

\begin{figure}
\vskip21.0pc
\special{hscale=65 vscale=65 hoffset=45 voffset=-10
psfile=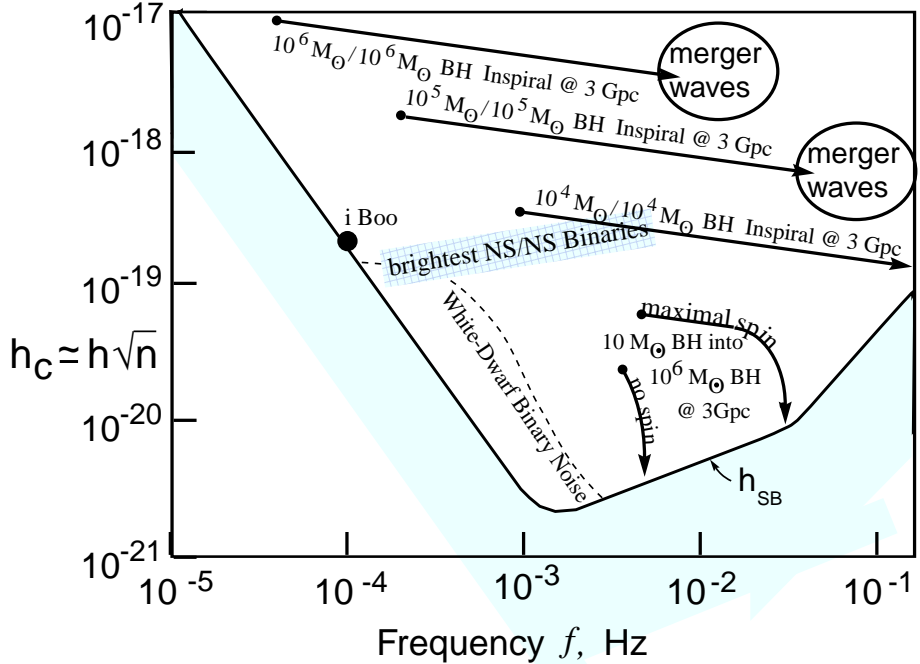}
\caption{LISA's projected sensitivity
to bursts $h_{\rm SB}$, compared with the strengths of the waves from
several low-frequency sources \protect\cite{cornerstone}. 
}
\label{fig:lisa_noise}
\end{figure}

Figure \ref{fig:lisa_noise} depicts the expected sensitivity of
LISA in the same language as we have used for LIGO (Fig.\
\ref{fig:ligosources}):  
$h_{\rm SB} =  5\sqrt5 h_{\rm rms}$ is the sensitivity
for high-confidence detection ($S/N=5$) of a signal coming from 
a random direction, assuming Gaussian noise.   

At frequencies $f\agt 10^{-3}$Hz, LISA's noise is due to photon counting
statistics (shot noise).  The sensitivity curve steepens at $f\sim
3\times10^{-2}$Hz because at larger $f$ than that, the waves' period is
shorter than the round-trip light travel time in one of LISA's arms. 
Below $10^{-3}$Hz, the noise is due to buffeting-induced random motions 
of the spacecraft
that are not being properly removed by the drag-compensation system.
Notice that, in terms of dimensionless amplitude, LISA's sensitivity is
roughly the same as that of LIGO's first interferometers (Fig.\ 
\ref{fig:ligosources}), but at 100,000 times lower frequency.  Since
the waves' energy flux scales as $f^2 h^2$, this corresponds to $10^{10}$
better energy sensitivity than LIGO.

LISA can detect and study, simultaneously, a wide variety of different
sources scattered over all directions on the sky.  The key to
distinguishing the different sources is the different time evolution of
their waveforms.  The key to determining each source's direction, and
confirming that it is real and not just noise, is the manner in which
its waves' amplitude and frequency are modulated by LISA's complicated
orbital motion---a motion in which the interferometer triangle rotates around
its center once per year, and the interferometer plane precesses 
around the normal to the Earth's orbit once per year.  Most sources will
be observed for a year or longer, thereby making full use of these
modulations.

\section{Stellar Core Collapse: The Births of Neutron Stars and Black Holes}

In the remainder of this article, I shall describe the techniques and prospects
for observationally studying black holes and relativistic stars via the
gravitational waves they emit.  I begin with the births of stellar-mass
neutron stars and black holes.

When the core of a massive star has exhausted its supply of nuclear fuel, 
it collapses to form a neutron star 
or a black hole. In some cases, the
collapse triggers and powers a subsequent explosion 
of the star's mantle---a supernova explosion.  Despite extensive
theoretical efforts for more than 30 years, and despite wonderful
observational data from Supernova 1987A, theorists are still far from  
a definitive understanding of the details of the collapse and explosion.  The
details are highly complex and may differ greatly from one
core collapse to another \cite{petschek}.  

Several features of the collapse and the core's subsequent
evolution can produce significant gravitational radiation in the 
high-frequency band. We shall
consider these features in turn, the most weakly radiating first, 
and we shall focus primarily on collapses that produce neutron stars rather 
than black holes.

\subsection{Boiling of a Newborn Neutron Star}

Even if the collapse is spherical, so it cannot radiate any
gravitational waves at all, it should
produce a convectively unstable neutron
star that ``boils'' vigorously (and nonspherically) for the first 
$\sim 1$ second of its life \cite{bethe}.  The boiling dredges 
up high-temperature  
nuclear matter ($T\sim 10^{12}$K) from the neutron star's central regions,
bringing it to the surface (to the ``neutrino-sphere''), where it 
cools by
neutrino emission before being swept back downward and reheated.  Burrows
\cite{burrows1} has pointed out that the  boiling
should generate $n \sim 100$ cycles of gravitational waves with
frequency $f\sim 100$Hz and amplitude large enough to be detectable by
LIGO/VIRGO throughout our galaxy and its satellites.
Neutrino
detectors have a similar range, and there could be a high scientific payoff
from
correlated observations of the gravitational waves emitted by the
boiling's mass motions and neutrinos emitted from the boiling
neutrino-sphere.  With neutrinos to trigger on, the sensitivities of LIGO
detectors should be about twice as good as shown in Fig.\
\ref{fig:ligosources}. 

Recent 3+1 dimensional simulations by 
M\"uller and Janka \cite{muller_janka} suggest an rms
amplitude
$h \sim 2 \times 10^{-23} (20{\rm kpc}/r)$ (where $r$ is the distance to
the source), corresponding to a characteristic amplitude $h_c \simeq
h\sqrt n \sim 2\times 10^{-22} (20{\rm kpc}/r)$; cf.\ Fig.\ 
\ref{fig:ligosources}.  (The older 2+1 dimensional simulations gave $h_c$ 
about 6 times larger than this \cite{muller_janka}, but presumably were less 
reliable.)  
LIGO should be 
able to detect such waves throughout our galaxy with an amplitude signal to 
noise ratio of about 
$S/N = 2.5$ in each of its two enhanced 4km interferometers, and its advanced
interferometers should do the same out to 80Mpc distance. 
(Recall that the $h_{\rm
SB}$ curves in Fig.\ \ref{fig:ligosources} are drawn at a signal to noise ratio
of about 5).  Although the estimated event rate is only about one every
40 years in our galaxy and not much larger out to 80Mpc, if just one such
supernova is detected the 
correlated neutrino and gravitational wave observations could bring very
interesting insights into the boiling of a newborn neutron star.

\subsection{Axisymmetric Collapse, Bounce, and Oscillations}

Rotation will centrifugally flatten the collapsing
core, enabling it to radiate as it implodes.  If the core's angular
momentum is small enough that centrifugal forces do 
not halt or strongly slow the collapse before it reaches
nuclear densities, then the core's collapse, bounce, and subsequent
oscillations are likely to be axially symmetric.  Numerical
simulations \cite{finn_collapse,monchmeyer} show that in this case the waves
from collapse, bounce, and oscillation 
will be quite weak: the total energy radiated as gravitational waves
is not likely to exceed $\sim 10^{-7}$ solar masses (about 1 part in a
million of the collapse energy) and might often be much less than
this; and correspondingly, the waves' 
characteristic amplitude will be $h_c \alt 3\times 10^{-21}(30{\rm
kpc}/r)$.  These collapse-and-bounce waves will come off at frequencies
$\sim 200$ Hz to $\sim 1000$ Hz, and will precede the boiling waves by a
fraction of a second.  Though a little stronger than the boiling waves, 
they probably cannot be seen by LIGO/VIRGO beyond the local group of 
galaxies and thus will be a very rare occurrence.

\subsection{Rotation-Induced Bars and Break-Up}

If the core's rotation is large enough to strongly flatten the 
core before or as it reaches nuclear density, 
then a dynamical or
secular instability is likely to break the core's axisymmetry.
The core will be transformed into
a bar-like configuration that spins end-over-end like
an American football, and that might even break up into two or more massive
pieces.  As we shall see below, the radiation from the spinning bar or orbiting
pieces {\it could} be almost as strong as that from a coalescing neutron-star
binary (Sec.\ \ref{nsnsinspiral}), 
and thus could be seen by the LIGO/VIRGO first 
interferometers
out to the distance of the Virgo cluster (where the supernova rate is 
several per
year), by enhanced interferometers out to $\sim 100$Mpc (supernova rate several
thousand per year), and by advanced interferometers out to several hundred 
Mpc (supernova rate $\sim \hbox{(a few)}\times 10^4$ per year); cf.\ 
Fig.\  \ref{fig:ligosources}.  It is far 
from clear what fraction of collapsing cores will have enough angular
momentum to break their axisymmetry, and what fraction of those will
actually radiate at this high rate; but even if only $\sim 1/1000$ or
$1/10^4$ do so, this could ultimately be a very interesting source for
LIGO/VIRGO.

Several specific scenarios for such non-axisymmetry have been identified:

{\bf Centrifugal hangup at $\bf \sim 100$km radius:} If the
pre-collapse core is rapidly spinning (e.g., if it is a white dwarf that
has been spun up by accretion from a companion), then the collapse may
produce a highly flattened, centrifugally supported disk with most of
its mass at radii $R\sim 100$km, which then (via instability)
may transform itself into a bar or may bifurcate.  The bar or
bifurcated lumps will radiate gravitational waves at twice their rotation 
frequency, $f\sim 100$Hz---the optimal frequency for LIGO/VIRGO
interferometers.  To shrink on down to $\sim 10$km size, this
configuration must shed most of its angular momentum.  {\it If} a
substantial fraction of the angular momentum goes into
gravitational waves, then independently of the strength of the bar,
the waves will be nearly as strong as those from a coalescing binary.
The reason is this:
The waves' amplitude $h$ is proportional to the bar's ellipticity $e$,
the number of cycles $n$ of wave emission is proportional to $1/e^2$, and the
characteristic amplitude $h_c = h\sqrt n$ is thus independent of the
ellipticity and is about the same whether the configuration is a bar or
is two lumps \cite{schutz_grg89}.  The resulting waves will thus have $h_c$
roughly half as large, at $f\sim 100$Hz, as the $h_c$ from a NS/NS binary
(half as large because each lump might be half as massive as a NS), and
the waves will chirp upward in frequency in a manner similar to those from a
binary (Sec.\ \ref{nsnsinspiral}).

It may very well be, however, that most of the core's
excess angular momentum does {\it
not} go into gravitational waves, but instead goes largely into hydrodynamic
waves as the bar or lumps, acting like a propeller, stir up the 
surrounding stellar mantle.  In this case, the radiation will be
correspondingly weaker. 

{\bf Centrifugal hangup at $\bf \sim 20$km radius:}  Lai and Shapiro
\cite{lai} have explored the case of centrifugal
hangup at radii not much larger than the final neutron star, say $R\sim
20$km.  Using compressible ellipsoidal models, they have deduced that,
after a brief period of dynamical bar-mode instability with wave
emission at $f\sim 1000$Hz (explored by
Houser, Centrella, and Smith \cite{houser}), the star switches to a secular
instability in which the bar's angular velocity gradually slows while
the material of which it is made retains its high rotation speed and
circulates through the slowing bar.  The slowing bar emits waves that sweep
{\it downward} in frequency through the LIGO/VIRGO optimal band $f\sim 100$Hz,
toward $\sim 10$Hz. The characteristic amplitude (Fig.\
\ref{fig:ligosources}) is only modestly smaller than for the upward-sweeping
waves from hangup at $R\sim 100$km, and thus such waves should be
detectable near the Virgo Cluster by the first LIGO/VIRGO interferometers,
near 100Mpc by enhanced interferometers, 
and at distances of a few 100Mpc by advanced interferometers.

{\bf Successive fragmentations of an accreting, newborn neutron star:}
Bonnell and Pringle \cite{pringle} have focused on the evolution of the
rapidly spinning, newborn neutron star as it quickly accretes more and
more mass from the pre-supernova star's inner mantle.  If the accreting
material carries high angular momentum, it may trigger a renewed bar
formation, lump formation, wave emission, and coalescence, followed by more
accretion, bar and lump formation, wave emission, and coalescence.  Bonnell
and Pringle
speculate that hydrodynamics, not wave emission, will drive this
evolution, but that the total energy going into gravitational waves might be
as large as $\sim 10^{-3}M_\odot$.  This corresponds to $h_c \sim 10^{-21}
(10{\rm Mpc}/r)$.

\section{Pulsars: Spinning Neutron Stars}
\label{pulsars}

As the neutron star settles down into its final state, its crust begins
to solidify (crystalize). The solid
crust will assume nearly the oblate axisymmetric shape that 
centrifugal forces are trying to maintain,
with poloidal 
ellipticity $\epsilon_p \propto$(angular velocity of rotation)$^2$. 
However, the principal axis
of the star's moment of inertia tensor may deviate from its spin axis
by some small ``wobble angle'' $\theta_w$, and the star may 
deviate slightly from axisymmetry about its principal axis; i.e., it may
have a slight ellipticity $\epsilon_e \ll \epsilon_p$ in its equatorial plane.

As this slightly imperfect crust spins, it will radiate gravitational
waves \cite{zimmermann}: $\epsilon_e$ radiates at twice the rotation 
frequency, $f=2f_{\rm rot}$ with
$h\propto \epsilon_e$, and the wobble angle couples to $\epsilon_p$ to
produce waves at $f=f_{\rm rot} + f_{\rm prec}$
(the precessional sideband of the rotation frequency) with amplitude
$h\propto \theta_w \epsilon_p$.  For typical neutron-star masses and
moments of inertia, the wave amplitudes are
\begin{equation}
h \sim 6\times 10^{-25} \left({f_{\rm rot}\over 500{\rm Hz}}\right)^2
\left({1{\rm kpc}\over r}\right)\left({\epsilon_e \hbox{ or }\theta_w\epsilon_p
\over 10^{-6}}\right)\;.
\label{hpulsar}
\end{equation}
  
The neutron star gradually spins down, due in part to gravitational-wave
emission but perhaps more strongly due to electromagnetic torques associated
with its spinning magnetic field and pulsar emission. 
This spin-down reduces the strength of centrifugal forces, and thereby
causes the star's poloidal ellipticity $\epsilon_p$ to decrease, with
an accompanying breakage and resolidification of its crust's crystal structure
(a ``starquake'') \cite{starquake}.  
In each starquake, $\theta_w$, $\epsilon_e$, and
$\epsilon_p$ will all change suddenly, thereby changing the amplitudes and
frequencies of the
star's two gravitational ``spectral lines'' $f=2f_{\rm rot}$ and
$f=f_{\rm rot} + f_{\rm prec}$.  After each quake, there should be a
healing period in which the star's fluid core and solid crust, now rotating
at different speeds, gradually regain synchronism.
By monitoring the 
amplitudes, frequencies, and phases of the two gravitational-wave
spectral lines, and by 
comparing with timing of
the electromagnetic pulsar emission, one might learn much about the 
physics of the neutron-star interior.

How large will be the quantities $\epsilon_e$ and $\theta_w \epsilon_p$?
Rough estimates of the crustal shear moduli and breaking strengths suggest an
upper limit in the range $\epsilon_{\rm max} \sim 10^{-4}$
to $10^{-6}$, and it might be that typical values are 
far below this.  We are extremely ignorant, and
correspondingly there is much to be learned from searches for
gravitational waves from spinning neutron stars.

One can estimate the sensitivity of LIGO/VIRGO (or any other broad-band
detector)
to the periodic waves from such a source by multiplying the waves'
amplitude $h$ by the square root of the number of cycles over which one
might integrate to find the signal, $n= f \hat \tau$ where $\hat\tau$ is the
integration time.  The resulting
effective signal strength, $h\sqrt{n}$, is larger than $h$ by
\begin{equation}
\sqrt n = \sqrt{f\hat\tau} = 10^5 \left( {f\over1000{\rm Hz}}\right)^{1/2}
\left({\hat\tau\over4{\rm months}}\right)^{1/2}\;.
\label{ftau}
\end{equation}
Four months of integration is not unreasonable in targeted searches; but for an
all-sky, all-frequency search, a coherent integration might not last longer
than a few days because of computational limitations associated with
having to apply huge numbers of trial neutron-star spindown corrections and
earth-motion doppler corrections \cite{brady_creighton_cutler_schutz}.

Equations (\ref{hpulsar}) and (\ref{ftau}) 
for $h\sqrt n$  should be compared (i) to the
detector's rms broad-band noise level for sources in a random direction,
$\sqrt5 h_{\rm rms}$, to deduce a
signal-to-noise ratio, or (ii) to $h_{\rm SB}$ to deduce a
sensitivity for
high-confidence detection when one does not know the waves' frequency in
advance \cite{300yrs}.   
Such a comparison suggests that the first interferometers in
LIGO/VIRGO might possibly see waves from nearby spinning
neutron stars, but the odds of success are very unclear.

The deepest searches for these nearly periodic waves will be
performed by narrow-band detectors, whose sensitivities are enhanced
near some chosen frequency at the price of sensitivity loss
elsewhere---signal-recycled interferometers \cite{meers}, 
resonant-sideband-extraction interferometers \cite{mizuno}, or
resonant-mass
antennas \cite{tiga,grail} (Section \ref{narrowband}).
With ``advanced-detector technology'' and targeted searches, such detectors
might be able to find with confidence spinning neutron stars 
that have \cite{300yrs}
\begin{equation}
(\epsilon_e \hbox{ or } \theta_w\epsilon_p ) \agt 3\times10^{-10} \left(
{500 {\rm Hz}\over f_{\rm rot}}\right)^2 \left({r\over 1000{\rm pc}}\right)^2.
\label{advancedpulsar}
\end{equation}
There may well be a large number of such neutron stars in our galaxy; but
it is also conceivable that there are none.  We are extremely
ignorant.

Some cause for optimism arises from several physical mechanisms that
might generate radiating ellipticities large compared to
$3\times10^{-10}$: 
\begin{itemize}

\item It may be that, inside the superconducting cores of
many neutron stars, there are trapped magnetic fields with mean
strength $B_{\rm core}\sim10^{13}$G or even
$10^{\rm 15}$G. 
Because such a field is actually concentrated in flux
tubes with $B = B_{\rm crit} \sim 6\times 10^{14}$G surrounded by
field-free superconductor, its mean pressure is $p_B = B_{\rm core} B_{\rm
crit}/8\pi$.  This pressure could produce a radiating 
ellipticity  
$\epsilon_{\rm e} \sim \theta_w\epsilon_p \sim p_B/p \sim 10^{-8}B_{\rm
core}/10^{13}$G (where $p$ is the core's material pressure). 

\item Accretion onto a spinning neutron star can drive precession (keeping
$\theta_w$ substantially nonzero), and thereby might produce measurably strong
waves \cite{schutz95}.

\item If a neutron star is born rotating very rapidly,
then it may experience a
gravitational-radiation-reaction-driven instability first discovered by
Chandrasekhar \cite{cfs_chandra} and elucidated in greater detail by 
Friedman and Schutz
\cite{cfs_friedman_schutz}).  In this ``CFS instability'',
density waves travel around the
star in the opposite direction to its rotation, but are dragged forward
by the rotation.  These density waves produce gravitational waves that 
carry positive energy as seen by observers far from the star, but
negative energy from the star's viewpoint; and because the
star thinks it is losing negative energy, its density waves get
amplified.  This intriguing mechanism is similar to that by which
spiral density waves are produced in galaxies.  Although the CFS
instability was once thought ubiquitous for spinning stars 
\cite{cfs_friedman_schutz,wagoner}, we now
know that neutron-star viscosity will kill it, stabilizing the star and
turning off the waves, when the star's temperature is above some
limit $\sim 10^{10}{\rm K}$ \cite{cfs_lindblom}
and below some limit $\sim 10^9 {\rm K}$
\cite{cfs_mendell_lindblom}; and correspondingly, the instability
should operate only during the first few years of a neutron
star's life, when $10^9 {\rm K} \alt T \alt 10^{10}\rm K$.

\end{itemize}

\section{Neutron-Star Binaries and Their Coalescence}
\label{nsbinaries}

\subsection{NS/NS and Other Compact Binaries in Our Galaxy}
\label{nsbinariesingalaxy}

The best understood of all gravitational-wave sources are binaries made of two
neutron stars (``NS/NS binaries'').
The famous Hulse-Taylor \cite{hulse_taylor,taylor} binary pulsar, 
PSR 1913+16, is an example. 
At present PSR 1913+16 has an orbital frequency of about 1/(8 hours)
and emits its waves predominantly at twice this frequency, roughly
$10^{-4}$ Hz, which is in LISA's low-frequency band (Fig.\
\ref{fig:lisa_noise}); but it is too weak for LISA to detect.  
LISA will be able to search for brighter NS/NS binaries in
our galaxy with periods shorter than this.  

If conservative estimates \cite{narayan,phinney,vdheuvellorimer} 
based on the statistics of binary pulsar observations are correct,
there should be many NS/NS binaries in our galaxy that are brighter in
gravitational waves than PSR 1913+16.  Those estimates suggest that
one compact NS/NS binary is born every $10^5$ years in our galaxy
and that the brightest NS/NS binaries will
fall in the indicated region in Fig.\ \ref{fig:lisa_noise}, extending out to a
high-frequency limit of $\simeq 3\times 10^{-3}$Hz (corresponding to a
remaining time to coalescence of $10^5$ years).  The
birth rate might be much higher than $1/10^5$years, 
according to progenitor evolutionary arguments 
\cite{phinney,tutukov_yungelson,yamaoka,lipunov,lipunov1,zwartspreeuw},
in which case LISA would
see brighter and higher-frequency binaries than shown in Fig.\
\ref{fig:lisa_noise}.  LISA's observations should easily reveal the true
compact NS/NS birth rate and also the birth rates of NS/BH and BH/BH 
binaries---classes of objects
that have not yet been discovered electromagnetically.  For further details see
\cite{wdbinaries,cornerstone}; 
for estimates of LISA's angular resolution when observing
such binaries, see \cite{cutlerlisa}.

\subsection{The Final Inspiral of a NS/NS Binary}
\label{nsnsinspiral}

As a result of their loss of orbital energy to gravitational waves,
the PSR 1913+16 NS's are gradually spiraling inward at a rate that agrees 
with general relativity's prediction to 
within the measurement accuracy (a fraction of a percent) 
\cite{taylor}---a remarkable but indirect confirmation that
gravitational waves do exist and are correctly described by general
relativity.  If we wait roughly
$10^8$ years, this inspiral will bring the waves into the LIGO/ VIRGO
high-frequency band. As the NS's continue their inspiral, over a time of about
15 minutes
the waves will sweep through the LIGO/VIRGO band, 
from $\sim 10$ Hz to $\sim 10^3$ Hz, at which point the NS's will
collide and merge.  It is this last 15 minutes 
of inspiral, with $\sim 16,000$
cycles of waveform oscillation, and the final merger,
that the LIGO/VIRGO network seeks to monitor.

To what distance must LIGO/VIRGO look, in order to see such inspirals several
times per year?
Beginning with our galaxy's conservative, pulsar-observation-based NS/NS event
rate of one every 100,000 years (Sec.\ \ref{nsbinariesingalaxy}) and
extrapolating out through the Universe, one infers an event rate of
several per year at 200 Mpc \cite{narayan,phinney,vdheuvellorimer}.  
If arguments based on simulations of
binary evolution are correct 
\cite{phinney,tutukov_yungelson,yamaoka,lipunov,lipunov1,zwartspreeuw}
(Sec.\ \ref{nsbinariesingalaxy}), the distance
for several per year could be as small as 23 Mpc---though such a small distance
entails stretching all the numbers to near their breaking point of 
plausibility \cite{phinney}.
If one stretches all numbers to the opposite, most pessimistic extreme, one
infers several per year at 1000 Mpc \cite{phinney}.  
Whatever may be the true distance for
several per year, once LIGO/VIRGO reaches that distance, each further
improvement of sensitivity by a factor 2 will increase the observed event rate
by $2^3 \simeq 10$.

Figure \ref{fig:ligosources} 
compares the projected LIGO sensitivities 
\cite{ligoscience} with the wave strengths from 
NS/NS inspirals at various distances from Earth.  From that comparison we see
that LIGO's first interferometers
can reach 30Mpc, where the most extremely optimistic estimates predict
several per year; the enhanced interferometers can reach 300Mpc where the
binary-pulsar-based, conservative estimates predict $\sim 10$ per year;
the advanced interferometers can reach 1000Mpc where even the most extremely
pessimistic of estimates predict several per year.

\subsection{Inspiral Waveforms and the Information they Carry}
\label{inspiralwaveforms}
 
Neutron stars have such intense self gravity that it is
exceedingly difficult to deform them.  Correspondingly, as they spiral
inward in a compact binary, they do not gravitationally deform each other
significantly until several orbits before their final
coalescence \cite {kochanek,bildsten_cutler}.  This means
that the inspiral waveforms are determined to high accuracy by 
only a few, clean parameters:
the masses and spin angular momenta of the stars, and the initial
orbital elements (i.e.\ the elements when the waves enter the LIGO/VIRGO band). 
The same is true for NS/BH and BH/BH binaries.  The following description of
inspiral waveforms is independent of whether the binary's bodies are NS's or
BH's.

\begin{figure}
\vskip 14.6pc
\special{hscale=65 vscale=65 hoffset=35 voffset=-13
psfile=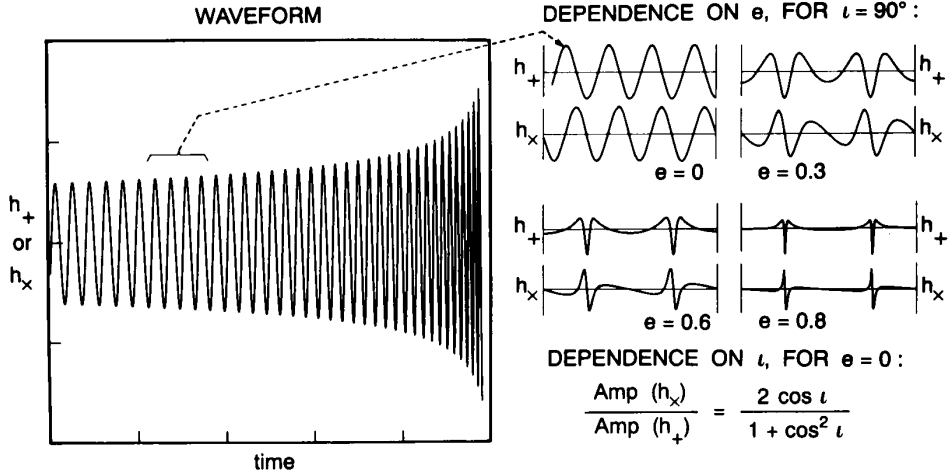}
\caption{Waveforms from the inspiral of a compact binary (NS/NS, NS/BH, or
BH/BH), computed using
Newtonian gravity for the orbital evolution and the quadrupole-moment
approximation for the wave generation.  (From Ref.\ 
\protect\cite{ligoscience}.)}
\label{fig:NewtonInspiral}
\end{figure}

Though tidal deformations are negligible during inspiral, relativistic
effects can be very important. 
If, for the moment, we ignore the relativistic effects---i.e., if we
approximate gravity as Newtonian and the wave generation as due to the
binary's oscillating quadrupole moment \cite{300yrs}, 
then the shapes of the inspiral
waveforms $h_+(t)$ and $h_\times(t)$
are as shown in Fig.\  \ref{fig:NewtonInspiral}.

The left-hand graph in Fig.\  \ref{fig:NewtonInspiral} shows the waveform 
increasing in
amplitude and sweeping upward in frequency 
(i.e., undergoing a ``chirp'') 
as the binary's bodies spiral closer and closer together.  The ratio of 
the amplitudes
of the two polarizations is determined by the inclination $\iota$ of the
orbit to our line of sight (lower right in Fig.\ \ref{fig:NewtonInspiral}).  The
shapes of the individual waves, i.e.\ the waves' harmonic content, are
determined by the orbital eccentricity (upper right).  (Binaries
produced by normal stellar evolution should be highly circular due to
past radiation reaction forces, but compact
binaries that form by capture events, in dense star clusters that might
reside in galactic nuclei \cite{quinlan_shapiro}, could be quite 
eccentric.)  If, for simplicity, the 
orbit is circular, then the rate at which 
the frequency sweeps or ``chirps'', $df/dt$ 
[or equivalently the number of cycles
spent near a given frequency, $n=f^2(df/dt)^{-1}$] is determined solely, in the
Newtonian/quadrupole approximation, by the binary's so-called {\it
chirp mass}, $M_c \equiv (M_1M_2)^{3/5}/(M_1+M_2)^{1/5}$ (where $M_1$ 
and $M_2$ are the two bodies' masses). 
The amplitudes of the two waveforms are determined by the chirp mass,
the distance to the source, and the orbital inclination.  Thus 
(in the Newtonian/quadrupole
approximation), by measuring the two amplitudes, the frequency sweep, and
the harmonic content of the inspiral waves, one can determine as direct,
resulting observables, the source's distance, chirp mass, inclination,
and eccentricity \cite{schutz_nature86,schutz_grg89}.
(For binaries at cosmological distances, the observables are the ``luminosity
distance,'' ``redshifted'' chirp mass $(1+z)M_c$, inclination, and
eccentricity; 
cf.\ Sec.\ \ref{bhbhstrength}.)

As in binary pulsar observations \cite{taylor}, 
so also here, relativistic effects add further information:
they influence the rate of frequency sweep and produce waveform
modulations in ways that
depend on the binary's dimensionless ratio $\eta = \mu/M$ of reduced mass 
$\mu = M_1 M_2/(M_1 + M_2)$ to total mass $M = M_1 + M_2$ 
and on the spins of the binary's two bodies. 
These relativistic effects are reviewed and discussed at length in 
Refs.\ \cite{last3minutes,will_nishinomiya}.  Two deserve special
mention: (i) As the waves emerge from the binary, some of them get
backscattered one or more times off the binary's spacetime curvature,
producing wave {\it tails}.  These tails act back on the binary,
modifying its radiation reaction force and thence its
inspiral rate in a measurable way.  (ii)  If the orbital 
plane is inclined to one or both of the binary's spins, then 
the spins drag inertial frames in the binary's
vicinity (the ``Lense-Thirring effect''), this frame dragging causes
the orbit to precess, and the precession modulates the 
waveforms \cite{last3minutes,precess,kidder}.  

Remarkably, the relativistic corrections to the frequency sweep---tails, 
spin-induced precession and others---will be 
measurable with rather high
accuracy, even though they are typically $\alt 10$ per cent of the 
Newtonian contribution, and even though the
typical signal to noise ratio will be only $\sim 9$. 
The reason is as 
follows \cite{cutler_flanagan,finn_chernoff,last3minutes}:

The frequency sweep will be monitored by the method of ``matched
filters''; in other words, the incoming, noisy signal will be cross 
correlated with theoretical templates.  If the signal and the templates
gradually 
get out of phase with each other by more than $\sim 1/10$ cycle as the 
waves sweep
through the LIGO/VIRGO band, their cross correlation will be significantly
reduced.
Since the total number of cycles spent in the LIGO/VIRGO band will
be $\sim 16,000$ for a NS/NS binary, $\sim 3500$ for NS/BH, and $\sim
600$ for BH/BH, this means that LIGO/VIRGO should be able to measure the
frequency sweep to a fractional precision $\alt 10^{-4}$, 
compared to which the relativistic effects
are very large.  (This is essentially the same method as
Joseph Taylor and colleagues use for high-accuracy radio-wave measurements of
relativistic effects in binary pulsars \cite{taylor}.)

Analyses using the theory of optimal signal processing
predict the following typical accuracies for LIGO/VIRGO measurements
based solely on the frequency sweep (i.e., ignoring modulational
information)
\cite{poisson_will}: (i) The chirp mass $M_c$
will typically be measured, from the Newtonian part of the frequency
sweep, to $\sim 0.04\%$ for a NS/NS binary and
$\sim 0.3\%$ for a system containing at least one BH.
(ii) {\it If} we
are confident (e.g., on a statistical basis from measurements of many
previous binaries) that the spins are a few percent or less
of the maximum physically allowed, then the reduced mass $\mu$
will be measured to
$\sim 1\%$ for NS/NS and NS/BH binaries, and
$\sim 3\%$ for BH/BH binaries.  (Here and below NS means a
$\sim 1.4 M_\odot$
neutron star and BH means a $\sim 10 M_\odot$
black hole.) (iii) Because the
frequency dependences
of the (relativistic) $\mu$ effects
and spin effects are not
sufficiently different
to give a clean separation between $\mu$ and the spins,
if we have no prior knowledge of the spins, then
the spin$/\mu$ correlation will
worsen the typical accuracy of $\mu$ by a large factor,
to $\sim 30\%$ for NS/NS, $\sim 50\%$ for NS/BH, and
a factor $\sim 2$ for BH/BH.
These worsened accuracies might be improved somewhat
by waveform modulations caused by the
spin-induced precession of the orbit \cite{precess,kidder},
and even without modulational information, a certain
combination of $\mu$ and the spins
will be determined to a few per cent.  Much
additional theoretical work is needed
to firm up the measurement accuracies.

To take full advantage of all the information in the inspiral waveforms
will require theoretical templates that are accurate, for given masses
and spins, to a fraction of a cycle during the entire sweep through the
LIGO/VIRGO band.  Such templates are being computed by an international
consortium of relativity theorists (Blanchet and Damour in France, Iyer
in India, Will and Wiseman in the U.S.)
\cite{2pnresults}, using post-Newtonian expansions of
the Einstein field equations, of the sort pioneered by Chandrasekhar 
\cite{chandracoll,chandra25}.  This enterprise is rather like computing
the Lamb shift to high order in powers of the fine structure
constant, for comparison with experiment and testing of quantum
electrodynamics.  Cutler
and Flanagan \cite{cutler_flanagan1} have estimated the order to which the
computations must be carried in order that
systematic errors in the theoretical templates will not significantly
impact the information extracted from the LIGO/VIRGO observational data.
The answer appears daunting: radiation-reaction effects must be computed
to three full post-Newtonian orders [six orders in $v/c =$(orbital
velocity)/(speed of light)] beyond Chandra's leading-order radiation reaction,
which itself is 5 orders in $v/c$ beyond the Newtonian theory of
gravity, so the required calculations are 
$O[(v/c)^{6+5}] = O[(v/c)^{11}]$.  By clever use of Pad\'e approximates, 
these requirements might be relaxed \cite{damour_sathya}.

In the late 1960's,
when Chandra and I were first embarking on our respective studies of
gravitational waves, Chandra set out to compute the first 5 orders in $v/c$
beyond Newton, i.e., in his own words, ``to solve Einstein's equations through
the 5/2 post-Newtonian'', thereby fully understanding leading-order radiation
reaction and all effects leading up to it.  
Some colleagues thought his project not worth the enormous
personal effort that he put into it.  But Chandra was prescient.  He had faith
in the importance of his effort, and history has proved him right.  The results
of his ``5/2 post-Newtonian'' \cite{chandra25}
calculation have now been verified to accuracy
better than 1\% by observations of the inspiral of PSR 1913+16; and the needs
of LIGO/VIRGO data analysis are now driving the calculations onward from 
$O[(v/c)^5]$ to $O[(v/c)^{11}]$.  This epitomizes
a major change in the field of relativity research: At last, 80 years
after Einstein formulated general relativity, experiment has become a
major driver for theoretical analyses.

Remarkably, the goal of $O[(v/c)^{11}]$ is achievable.  The most difficult
part of the computation, the radiation reaction, has been evaluated to
$O[(v/c)^9]$ beyond Newton by the French/Indian/ American consortium
\cite{2pnresults} 
and $O[(v/c)^{11}]$ is now being pursued.

These high-accuracy waveforms are needed only for extracting information
from the inspiral waves after the waves have been discovered; they are
not needed for the discovery itself.  The discovery is best achieved
using a different family of theoretical waveform templates, one that 
covers the space of potential waveforms
in a manner that minimizes computation time instead
of a manner that ties quantitatively into general relativity 
theory \cite{last3minutes,owen}.  Such templates are under development.

\subsection{NS/NS Merger Waveforms and their Information}
\label{merger_waves}

The final merger of a NS/NS binary should produce waves that are
sensitive to the equation of state of nuclear matter, so
such mergers have the potential to teach us about the
nuclear equation of state \cite{ligoscience,last3minutes}.  In essence, 
LIGO/VIRGO will be 
studying nuclear physics via the collisions of atomic nuclei that have
nucleon numbers $A \sim 10^{57}$---somewhat larger than physicists are normally 
accustomed to. 
The accelerator used to drive these ``nuclei'' up to half the speed of light is
the binary's self gravity, and the radiation by which the details of the
collisions are probed is gravitational.

Unfortunately, the NS/NS merger will emit its gravitational
waves in the kHz frequency band ($600 {\rm Hz} \alt f \alt 2500 {\rm
Hz}$) where photon shot noise will prevent the waves from being studied by
the standard, ``workhorse,'' broad-band
interferometers of Fig.\  \ref{fig:ligosources}.
However, it may be possible to measure the waves and extract their
equation-of-state information using
a ``xylophone'' of specially configured narrow-band detectors 
(signal-recycled or
resonant-sideband-extraction interferometers, and/or spherical or icosahedral 
resonant-mass detectors; Sec.\ \ref{narrowband} and Refs.\ 
\cite{last3minutes,kennefick_laurence_thorne}). Such measurements will
be very difficult and are likely only when the LIGO/VIRGO
network has reached a mature stage.  

A number of research groups \cite{centrella}
are engaged in numerical 
simulations of NS/NS mergers, with the goal not only to predict the
emitted gravitational waveforms and their dependence on equation of
state, but also (more immediately) to learn whether such 
mergers
might power the $\gamma$-ray bursts that have been a major astronomical
puzzle since their discovery in the early 1970s.  

NS/NS mergers are a
promising explanation for $\gamma$-ray bursts because 
(i) some bursts are known, from intergalactic absorption lines, to come from 
cosmological distances \cite{metzger}, (ii) the bursts have
a distribution of number versus intensity that suggests most lie
at near-cosmological distances, 
(iii) their event rate is
roughly the same as that conservatively estimated for NS/NS mergers 
($\sim1000$ per year out to cosmological distances; a few per year at 
300Mpc);
and (iv) it is plausible that the final NS/NS merger will create a 
$\gamma$-emitting fireball with enough energy to account for the bursts
\cite{meszaros,woosley}.
If enhanced LIGO interferometers were now in operation and observing
NS/NS inspirals, they could
report definitively whether or not the $\gamma$-bursts are produced by
NS/NS binaries; and if the answer were yes, then the combination of
$\gamma$-burst data and gravitational-wave data could bring valuable
information that neither could bring by itself.  For example, it would
reveal when, to within a few msec, the $\gamma$-burst is emitted 
relative to the moment the NS's first begin to touch; and by
comparing the $\gamma$ and gravitational times of arrival, 
we could test whether gravitational waves propagate with
the speed of light to a fractional precision of 
$\sim 0.01{\rm sec}/10^9\, {\rm lyr} = 3\times 10^{-19}$.

\subsection{NS/BH Mergers}

A neutron star (NS) spiraling into a black hole of mass $M \agt 10
M_\odot$ should be swallowed more or less whole.  However, if the BH is
less massive than roughly $10 M_\odot$, and especially if it is rapidly
rotating, then the NS will tidally disrupt before being swallowed.
Little is known about the disruption and accompanying waveforms.  To
model them with any reliability will likely require full numerical
relativity, since the circumferences of the BH and NS will be comparable
and their physical separation at the moment of disruption
will be of order their separation. As with NS/NS, the merger 
waves should
carry equation of state information and will come out in the kHz band,
where their detection will require advanced, specialty detectors.

\section{Black Hole Binaries}
\label{bhbh}

\subsection{BH/BH Inspiral, Merger, and Ringdown}
\label{bhbhinspmergring}

We turn, next, to binaries made of two black holes with comparable masses
(BH/BH binaries).  The LIGO/VIRGO network can detect and study waves from
the last few minutes of the life of such a binary if its total mass is $M\alt
1000 M_\odot$ (``stellar-mass black holes''), 
cf.\ Fig.\ \ref{fig:ligosources}; and LISA can do the same 
for the mass range $1000 M_\odot \alt M \alt 10^8 M_\odot$ (``supermassive
black holes''), cf. Fig.\ \ref{fig:lisa_noise}.  

The timescales for the binary's dynamics and its waveforms are
proportional to its total mass $M$. All other aspects of the 
dynamics and waveforms, after time scaling, depend solely on 
quantities that are dimensionless in geometrized units ($G=c=1$): the ratio 
of the 
two BH masses, the BH spins divided by the squares of their masses, etc.
Consequently, the black-hole physics to be studied is the same for supermassive
holes in LISA's low-frequency band as for stellar-mass holes in LIGO/VIRGO's 
high-frequency band.  LIGO/VIRGO is likely to make
moderate-accuracy studies of this physics; and LISA, flying later, can 
achieve high accuracy.

The binary's dynamics and its emitted waveforms can be divided into three
epochs: {\it inspiral, merger, and ringdown} \cite{hughes_flanagan}.  
The inspiral epoch terminates
when the holes reach their last stable orbit and begin plunging toward each
other.  The merger epoch lasts from the beginning of plunge until the holes
have merged and can be regarded as a single hole undergoing 
large-amplitude, quasinormal-mode vibrations. 
In the ringdown epoch,
the hole's vibrations decay due to wave emission, leaving finally a
quiescent, spinning black hole.

The inspiral epoch has been well studied theoretically using post-Newtonian
expansions (Sec.\ \ref{inspiralwaveforms}), 
except for the last factor $\sim 3$ of upward
frequency sweep, during which the post-Newtonian expansions may fail.  The
challenge of computing this last piece of the inspiral is called the
``intermediate binary black hole problem'' (IBBH) and is a subject of 
current research in my own group and elsewhere.  
The merger epoch can be studied theoretically only via supercomputer 
simulations.  Techniques for such simulations
are being developed by several research groups, including an
eight-university American consortium of numerical relativists and
computer scientists called the
Binary Black Hole Grand Challenge Alliance \cite{GC}.  Chandrasekhar and
Detweiler \cite{chandradetweiler,detweiler} pioneered the study of the 
ringdown epoch
using the Teukolsky formalism for first-order perturbations of spinning (Kerr)
black holes (see Chandra's classic book \cite{chandrabh}), and the ringdown 
is now rather
well understood except for the strengths of excitation of the various
vibrational modes, which the merger observations and computations should
reveal.

The merger epoch, as yet, is very poorly understood. We can expect it to
consist of
large-amplitude, highly nonlinear vibrations of spacetime
curvature---a phenomenon of
which we have very little theoretical understanding today.  Especially
fascinating will be the case of two spinning black holes whose spins are
not aligned with each other or with the orbital angular momentum.  Each
of the three angular momentum vectors (two spins, one orbital) will drag
space in its vicinity into a tornado-like swirling motion---the general
relativistic ``dragging of inertial frames''---so the binary is rather
like two tornados with orientations skewed to each other, embedded inside a 
third, larger tornado with a third orientation.  The dynamical evolution of 
such a complex configuration of coalescing spacetime warpage, 
as revealed by its
emitted waves, might bring us
surprising new insights into relativistic gravity \cite{ligoscience}.  

\subsection{BH/BH Signal Strengths and Detectability}
\label{bhbhstrength}

\begin{figure}
\vskip 18.6pc
\special{hscale=50 vscale=50 hoffset=75 voffset=0
psfile=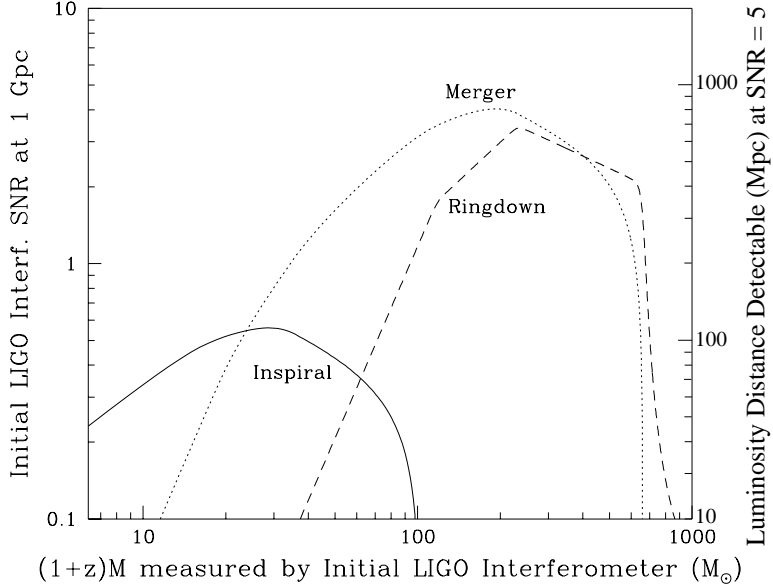}
\caption{The inspiral, merger, and ringdown waves from equal-mass black-hole
binaries as observed by LIGO's initial interferometers: The luminosity
distance to which
the waves are detectable (right axis) and the signal-to-noise ratio for a
binary at 1Gpc (left axis), as functions of the binary's redshifted
total mass (bottom axis).  (Figure adapted from Flanagan and Hughes
\protect\cite{hughes_flanagan}.)
}
\label{fig:ligo1bh}
\end{figure}

\begin{figure}
\vskip 22.6pc
\special{hscale=50 vscale=50 hoffset=70 voffset=0
psfile=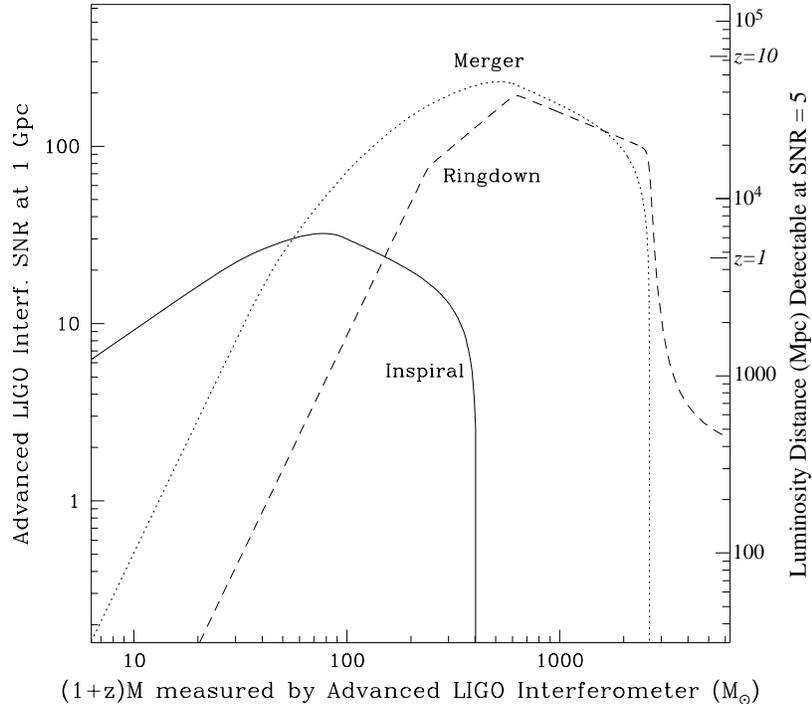}
\caption{The waves from equal-mass black-hole
binaries as observed by LIGO's advanced interferometers; cf.\ the caption for
Fig.\ \protect\ref{fig:ligo1bh}.  On the right side is
shown not only the luminosity distance to which the signals can be seen
(valid for any homogeneous, isotropic cosmology), 
but also the corresponding cosmological redshift $z$, 
assuming vanishing cosmological constant, a spatially flat universe,
and a Hubble constant $H_o = 75$ km/s/Mpc. 
(Figure adapted from Flanagan and Hughes \protect\cite{hughes_flanagan}.)
}
\label{fig:ligoadvbh}
\end{figure}

\begin{figure}
\vskip 22.6pc
\special{hscale=50 vscale=50 hoffset=70 voffset=0
psfile=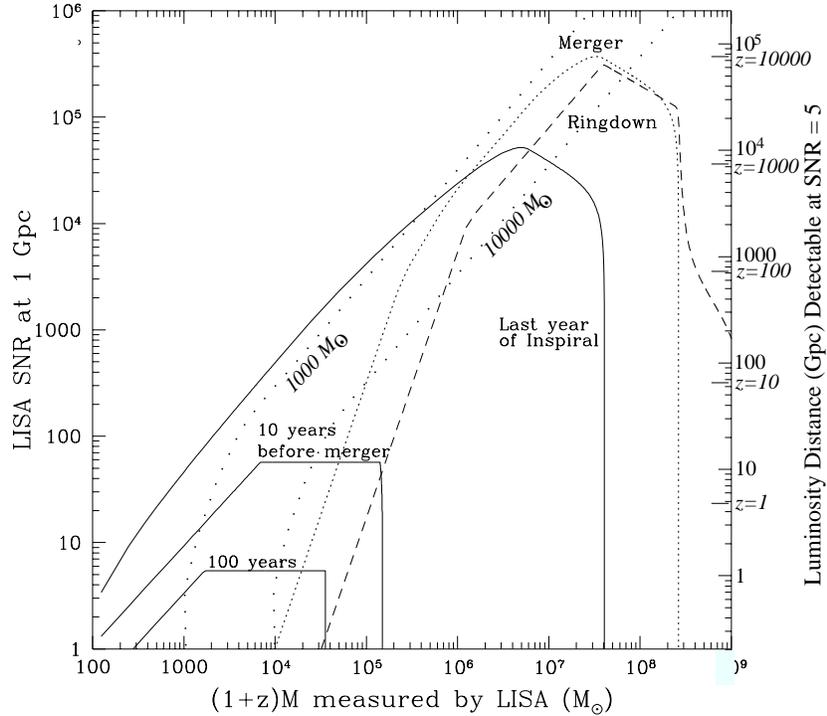}
\caption{The waves from equal-mass, supermassive black-hole
binaries as observed by LISA in one year of integration time; 
cf.\ the captions for 
Figs.\ \protect\ref{fig:ligo1bh} and \protect\ref{fig:ligoadvbh}.
The wide-spaced dots are curves of constant binary mass $M$, for use with the
right axis, assuming  
vanishing cosmological constant, a spatially flat universe,
and a Hubble constant $H_o = 75$ km/s/Mpc.  The bottom-most curves are the
signal strengths after one year of signal integration, for BH/BH binaries 10
years and 100 years before their merger.
(Figure adapted from Flanagan and Hughes \protect\cite{hughes_flanagan}.)
}
\label{fig:lisabh}
\end{figure}

Flanagan and Hughes \cite{hughes_flanagan} have recently
estimated the signal strengths
produced in LIGO and in LISA by the waves from equal-mass BH/BH binaries 
for each of the three epochs, 
inspiral, merger, and ringdown; and along with signal
strengths, they have estimated the distances to which LIGO and LISA can detect 
the waves.  In their 
estimates, Flanagan and Hughes 
make plausible assumptions about the waves' unknown aspects.  The estimated
signal strengths are shown in Fig.\ \ref{fig:ligo1bh} for the first LIGO
interferometers, Fig.\ \ref{fig:ligoadvbh} for advanced LIGO interferometers,
and Fig.\ \ref{fig:lisabh} for LISA.  Because LIGO and LISA can both reach out
to cosmological distances, these figures are drawn in a manner that includes
cosmological effects:  they are valid for any homogeneous,
isotropic model of our universe.  This is achieved by plotting 
observables that are extracted from the measured waveforms:
the binary's ``redshifted'' total mass $(1+z)M$ on the horizontal axis  
(where $z$ is the source's
cosmological redshift) and its
``luminosity distance'' \cite{luminosity_distance} on the right axis.  
The signal-to-noise ratio
(left axis) scales inversely with the luminosity distance.

We have no good observational handle on the coalescence rate of 
stellar-mass BH/BH binaries.  However, for BH/BH binaries with total mass
$M\sim 5$ to 50$M_\odot$ that arise from ordinary main-sequence 
progenitors,
estimates based on the progenitors' birth rates 
and on simulations of their 
subsequent evolution suggest a coalescence rate in our galaxy of
one per (1 to 30) million years \cite{lipunov1,tutukov_yungelson}.
These rough estimates 
imply that to see one coalescence per year with $M\sim 5$ to
$50M_\odot$, LIGO/VIRGO must reach out
to a distance $\sim ($300 to 900) Mpc.  Other plausible
scenarios (e.g. BH/BH binary
formation in dense stellar clusters that reside in globular clusters and 
galactic nuclei \cite{quinlan_shapiro}) 
could produce higher event rates and larger
masses, but little reliable is known about them (cf.\ Sec.\ I.A.ii of
\cite{hughes_flanagan}).

For comparison, the first LIGO interferometers can reach 300Mpc for
$M=50M_\odot$ but only 40Mpc for $M=5M_\odot$ (Fig.\ \ref{fig:ligo1bh}); 
enhanced interferometers can reach about 10 times farther, and advanced
interferometers about 30 times farther (Fig.\ \ref{fig:ligoadvbh}.  
These numbers suggest that (i) if waves from BH/BH coalescences 
are not detected 
by the first LIGO/VIRGO interferometers,
they are likely to be detected 
along the way from the first interferometers to the enhanced; and (ii) BH/BH
coalescences might be detected sooner than NS/NS coalescences (cf.\
Sec.\ \ref{nsnsinspiral}).  

For binaries with $M(1+z) \agt 40M_\odot$, the highly interesting merger 
signal should be stronger than the inspiral signal, and for $M\agt 100M_\odot$,
the ringdown should be stronger than inspiral (Fig.\ \ref{fig:ligo1bh}).  
Thus, it may well be that
early in the life of the LIGO/VIRGO network, observers and theorists
will be struggling to understand the merger of binary black holes by
comparison of computed and observed waveforms.  

LIGO's advanced interferometers 
(Fig.\ \ref{fig:ligoadvbh})  
can see the merger waves, for $20M_\odot \alt M
\alt 200M_\odot$) out to a cosmological redshift $z\simeq 5$; and for binaries
at $z=1$ in this mass range, they can achieve a signal to noise ratio (assuming
optimal signal processing \cite{hughes_flanagan}) of about 25 in each
interferometer.

While these numbers are impressive, they pale by comparison with LISA (Fig.\
\ref{fig:lisabh}), which can detect the merger waves for $1000M_\odot \alt M
\alt 10^5 M_\odot$ out to redshifts $z\sim 3000$ (far earlier in
the life of the universe than 
the era when the first supermassive black holes are likely to have formed). 
Correspondingly, LISA
can achieve signal to noise ratios of thousands for mergers 
with $10^5 \alt M \alt 10^8 M_\odot$ at redshifts of
order unity, and from the inspiral waves can infer the binary's parameters 
(redshifted
masses, luminosity distance, direction, ...) 
with high accuracy \cite{cutlerlisa}. 

Unfortunately, it is far from obvious whether the event rate for such 
supermassive
BH/BH coalescences will be interestingly high.  Conservative estimates suggest
a rate of $\sim 0.1/$yr, while plausible scenarios for aspects of the
universe about which we are rather ignorant 
can give rates as high as 1000/yr \cite{haehnelt}. 

If the coalescence rate is only 0.1/yr, then LISA should still see $\sim 3$ 
BH/BH
binaries with $3000M_\odot \alt M \alt 10^5 M_\odot$
that are $\sim 30$ years away from their final merger.  These slowly
inspiraling binaries should be visible, with one year of integration, out to 
a redshift $z \sim 1$ (bottom part of Fig.\ \ref{fig:lisabh}).

\section{Payoffs from Binary Coalescence Observations}
\label{payoffs}

Among the scientific payoffs that should come from LIGO/VIRGO's and/or LISA's
observations of binary coalescence are the following; others have been
discussed above.
 
\subsection{ Christodoulou Memory} 
\label{christodouloumemory}

As the gravitational waves from a binary's coalescence 
depart from
their source, the waves' energy creates (via the nonlinearity of Einstein's
field equations) a secondary wave called the ``Christodoulou memory''
\cite{christodoulou,thorne_memory,wiseman_will_memory}.  This memory, arriving
at Earth, can be
regarded rigorously as the combined gravitational field 
of all the gravitons
that have been emitted in directions other than toward the Earth
\cite{thorne_memory}. 
The memory builds up on the
timescale of the primary energy emission profile, and grows most rapidly 
when the primary waves are being emitted most strongly: during the end of 
inspiral and the merger.
Unfortunately, the memory is so weak that in LIGO only advanced
interferometers have much chance of detecting and studying 
it---and
then, only for BH/BH coalescences and not for NS/NS 
\cite{kennefick_memory}.  LISA, by contrast, should easily be able to measure
the memory from supermassive BH/BH coalescences.

\subsection{Testing General Relativity} 
\label{exotic_objects}

Corresponding to the very high post-Newtonian order to which a binary's
inspiral waveforms must be computed 
for use in LIGO/VIRGO and LISA data analysis
(Sec.\ \ref{inspiralwaveforms}), 
measurements of the inspiral waveforms can be used to test
general relativity with very high accuracy.  For example, in
scalar-tensor theories (some of which are attractive alternatives
to general relativity \cite{damour_nordvedt}), radiation reaction 
due to emission
of scalar waves places a unique signature on the measured inspiral
waveforms---a signature that can be searched for with high precision
\cite{will_scalartensor}.  Similarly, the inspiral waveforms can be used to
measure with high accuracy several fascinating general relativistic phenomena
in addition to the Christodoulou memory:
the influence of
the tails of the emitted waves on radiation reaction in the binary
(Sec.\ \ref{inspiralwaveforms}), the
Lens-Thirring orbital precision induced by the binary's spins
(Sec.\ \ref{inspiralwaveforms}), and a
unique relationship among the multipole moments of a quiescent black hole
which is dictated by a hole's ``two-hair theorem''
(Sec.\ \ref{lfbhinspiral}).

The ultimate test of general relativity will be detailed comparisons of the
predicted and observed waveforms from the highly nonlinear 
spacetime-warpage vibrations of BH/BH mergers (Sec.\ \ref{bhbhinspmergring}). 

\subsection{Cosmological Measurements}

Binary inspiral waves 
can be used to measure the Universe's Hubble constant, deceleration
parameter, and cosmological constant 
\cite{schutz_nature86,schutz_grg89,markovic,chernoff_finn}.  The keys to
such measurements are that: (i) Advanced interferometers in
LIGO/VIRGO will be able to see NS/NS inspirals
out to cosmological redshifts $z \sim 0.3$, and NS/BH out to $z
\sim 2$. (ii) The direct observables that can be extracted
from the
observed waveforms include a source's luminosity distance (measured
to an accuracy $\sim 10$ per cent in a large fraction of cases), and its
direction on the sky (to accuracy $\sim 1$ square degree)---accuracies
good enough that only one or a few electromagnetically-observed
clusters of galaxies should fall within the 3-dimensional
gravitational error boxes. This should make possible joint
gravitational/electromagnetic statistical studies of our Universe's 
magnitude-redshift 
relation, with gravity giving
luminosity distances and electromagnetism giving 
the redshifts \cite{schutz_nature86,schutz_grg89}.  (iii) Another direct
gravitational observable is any redshifted mass $(1+z)M$
in the system. Since the masses of NS's in binaries seem to
cluster around $1.4 M_\odot$, measurements of $(1+z)M$ can provide a
handle on the redshift, even in the absence of electromagnetic aid; so
gravitational-wave observations alone may be used, in a statistical way, to 
measure the magnitude-redshift relation \cite{markovic,chernoff_finn}.

LISA, with its ability to detect BH/BH binaries with $M\sim 1000$ to $100,000
M_\odot$ out to redshifts of
thousands, could search for the earliest epochs of supermassive black hole
activity---if the Universe is kind enough to grant us a large event rate. 

\subsection{Mapping Quiescent Black Holes; Searching for Exotic Relativistic
Bodies}
\label{lfbhinspiral}

Ryan \cite{ryanmap} has shown that,
when a white dwarf, neutron star or small black hole spirals into a 
much more massive, 
compact central body, the inspiral waves will carry a ``map'' of the 
massive body's external spacetime geometry.  Since the body's spacetime 
geometry is
uniquely characterized by the values of the body's multiple moments, we can say
equivalently that the inspiral waves carry, encoded in themselves, the values 
of all the body's multipole moments.  

By measuring the inspiral waveforms and extracting their map (i.e., measuring
the lowest few multipole moments), we can determine whether
the massive central body is a black hole or some other kind 
of exotic compact object \cite{ryanmap}; see below. 

The inspiraling object's orbital energy $E$ at fixed frequency
$f$ (and correspondingly at fixed orbital radius $a$)
scales as $E \propto \mu$, where $\mu$ is the object's mass; 
the gravitational-wave luminosity $\dot E$ scales as 
$\dot E \propto \mu^2$; and the time to
final merger thus scales as $t \sim E/\dot E \propto 1/\mu$.  This   
means that the smaller is $\mu/M$ (where $M$ is the central body's mass),
the more orbits are spent in the central body's strong-gravity region, $a\alt
10GM/c^2$, and thus the more detailed and accurate will be the map of the
body's spacetime geometry encoded in the emitted waves.

For holes observed by LIGO/VIRGO, the most extreme mass ratio that we
can hope for is $\mu/M \sim 1M_\odot/300 M_\odot$, since for $M>300M_\odot$ the
inspiral waves are pushed to frequencies below the LIGO/VIRGO band.  
This limit on $\mu/M$ seriously constrains the accuracy with which
LIGO/VIRGO can hope to map the spacetime geometry.
A detailed 
study by Ryan \cite{ryan_accuracy} (but one that is rather approximate because
we do not know the full details of the waveforms) 
suggests that LIGO/VIRGO might
{\it not} be able to distinguish cleanly between quiescent black holes and 
other types of massive central bodies.  

By contrast, LISA can observe the final inspiral waves from objects of
any mass $\mu\agt 1M_\odot$ spiraling into central bodies of mass 
$3\times 10^5 M_\odot \alt M \alt 3\times10^7M_\odot$ out to 3Gpc.  
Figure \ref{fig:lisa_noise} shows the
example of a $10M_\odot$ black hole spiraling into a $10^6M_\odot$ black hole
at 3Gpc distance.  The inspiral orbit and waves are strongly influenced
by the hole's spin.  Two cases are shown \cite{finn_thorne}: 
an inspiraling circular orbit 
around a non-spinning hole, and a prograde, circular, equatorial orbit 
around a maximally spinning hole. 
In each case the dot at the upper left end of the
arrowed curve is the frequency and characteristic amplitude one year
before the final coalescence.  In the nonspinning case, the small hole
spends its last year spiraling inward from $r\simeq 7.4 GM/c^2$ 
(3.7 Schwarzschild
radii) to its last stable circular orbit at $r=6GM/c^2$ (3 Schwarzschild
radii).  In the maximal spin case, the last year is spent traveling from
$r=6GM/c^2$ (3 Schwarzschild radii) to the last stable orbit at $r=GM/c^2$ 
(half a
Schwarzschild radius).  The $\sim 10^5$ cycles of waves during this last
year should carry, encoded in themselves, rather accurate values for 
the massive hole's lowest few multipole moments \cite{ryanmap,ryan_accuracy} 
(or, equivalently,
a rather accurate map of the hole's spacetime geometry).  

If the
measured moments satisfy the black-hole ``two-hair'' theorem 
(usually incorrectly called the ``no-hair'' theorem), i.e.\ 
if they are all
determined uniquely by the measured mass and spin in the manner of the
Kerr metric, then we can be sure the central body is a black hole.  If
they violate the two-hair theorem, then (assuming general relativity is
correct), either the central body was an exotic object---e.g. 
a spinning boson star which should have three ``hairs'' \cite{ryan_bosonstar},
a soliton star \cite{lee_pang} or a
naked singularity---rather than a black hole, or else an accretion
disk or other material was perturbing its orbit \cite{chakrabarti}. 
From the evolution of the waves one can hope to determine which is
the case, and to explore the properties of the central body and its
environment \cite{ryan_finn_thorne}.

Models of galactic nuclei, where massive holes (or other massive central
bodies) reside, suggest that
inspiraling stars and small holes typically will be in rather eccentric
orbits \cite{hils_bender,sigurdsson_rees}.  
This is because they get injected into such orbits via
gravitational deflections off other stars, and by the time gravitational
radiation reaction becomes the dominant orbital driving force, there is
not enough inspiral left to strongly circularize their orbits.  Such orbital
eccentricity will complicate the waveforms and complicate the extraction
of information from them.  Efforts to understand the emitted waveforms,
for central bodies with arbitrary multipole moments,
are just now getting underway \cite{ryanmap,ryan_waveforms}.  
Even for central black
holes, those efforts are at an early stage; for example, only recently have we
learned how to compute the influence of radiation reaction on inspiraling
objects in fully relativistic, nonequatorial orbits around a black hole
\cite{mino,wald}.

The event rates for inspiral into supermassive black holes (or other 
supermassive central bodies) are not 
well understood.  However, since a significant fraction of all galactic
nuclei are thought to contain supermassive holes, and since white dwarfs and
neutron stars, as well as small black holes, can withstand tidal 
disruption as they plunge toward a supermassive hole's horizon, and since
LISA can see inspiraling bodies as small as $\sim 1 M_\odot$ out to 
3Gpc distance, the event rate is likely to be interestingly large.  Sigurdsson
and Rees give a ``very conservative'' estimate of one inspiral event per year
within 1Gpc distance, and 100--1000 sources detectable by LISA at lower
frequencies ``en route'' toward their final plunge.

\section{Conclusion}
\label{conclusion}

It is now 37 years since Joseph Weber initiated his pioneering
development of gravitational-wave detectors \cite{weber}, 26 years
since Robert Forward \cite{forward} and Rainer Weiss \cite{weiss}
initiated work on interferometric detectors, and about 35 years since Chandra 
and others launched the modern era of theoretical research on
relativistic stars and black holes. Since then, hundreds of
talented experimental physicists have struggled to improve the
sensitivities of gravitational-wave detectors, and hundreds of theorists have
explored general relativity's predictions for stars and black holes.  

These two parallel efforts are now intimately intertwined and are pushing
toward an era in the not distant future, when measured gravitational waveforms
will be compared with theoretical predictions to learn how many and what kinds
of relativistic objects {\it really} populate our Universe, and how these
relativistic objects 
{\it really} are structured and {\it really}
behave when quiescent, when vibrating, and when colliding.

\section{Acknowledgments}
\label{acknowledgments}

My group's research on gravitational waves 
and their relevance to LIGO/VIRGO and LISA is supported in part
by NSF grants AST-9417371 and PHY-9424337 and by NASA grant 
NAGW-4268/NAG5-4351.
Large portions of this article were adapted and updated from my Ref.\ 
\cite{snowmass}. 


%
\end{document}